%% file: main.tex
\documentclass[10pt,conference]{IEEEtran}
\IEEEoverridecommandlockouts
\usepackage[utf8]{inputenc}
\usepackage[T1]{fontenc}
\usepackage[english]{babel}
\usepackage{csquotes}
\usepackage[style=ieee,sortcites=true,sorting=none, backend=biber,mincitenames=1, dashed=false, url = false,  doi=true, eprint= false, natbib=true, sorting=none, maxcitenames=1]{biblatex}

\usepackage{graphicx}
\usepackage{amsmath}
\usepackage{amssymb}
\usepackage{amsthm}
\usepackage{mathtools}
\usepackage{braket}
\usepackage{hyperref}
\usepackage{titlesec}
\usepackage{balance}
\usepackage{booktabs}
\usepackage{tabularx}
\usepackage{float}
\usepackage{tikz}
\usepackage{fontawesome}
\usepackage[skip=2pt]{caption}
\usepackage{subcaption}
\usepackage{fancyhdr}
\usepackage{float}
\usepackage{placeins}
\usepackage{cleveref}
\Crefname{figure}{Fig.}{Figs.}
\usepackage{algorithm}
\usepackage{algpseudocode}
\usepackage{dblfloatfix}

\graphicspath{{plots}}
\addto\extrasenglish{

}

\usepackage{fancyhdr}
 
\fancypagestyle{specialfooter}{%
  \fancyhf{}
  \renewcommand\headrulewidth{0pt}
  \fancyfoot[R]{ \noindent\fbox{%
    \parbox{\textwidth}{%
        {\footnotesize \copyright 2023 IEEE. Personal use of this material is permitted. Permission from IEEE must be obtained for all other uses, in any current or future media, including reprinting/republishing this material for advertising or promotional purposes, creating new collective works, for resale or redistribution to servers or lists, or reuse of any copyrighted component of this work in other works.}
        }
    }}
}

\title{Discovering Data Encoding Strategies for Quantum-Classical Neural Networks Using Monte Carlo Tree Search  \thanks{The project/research is supported by the Bavarian Ministry of Economic Affairs, Regional Development and Energy with funds from the Hightech Agenda Bayern.}}

\author{
    \IEEEauthorblockN{Lena Tokuhiro, Amine Bentellis, Jeanette Miriam Lorenz}
    \IEEEauthorblockA{Fraunhofer Institute for Cognitive Systems IKS\\
    \{lena.tokuhiro, amine.bentellis, jeanette.miriam.lorenz\}@iks.fraunhofer.de}
}

\begin{document}
\maketitle

\begin{abstract}
Quantum machine learning (QML) has attracted considerable research interest, yet whether it offers practical benefits over classical approaches remains an open question. The choice of data encoding significantly influences QML performance, but why certain encodings outperform others remains poorly understood.
We employ Monte Carlo Tree Search (MCTS) to discover optimal data encoding circuits for a quantum-classical convolutional neural network (QCCNN) combining a non-variational quantum block for feature extraction with a classical classifier. Evaluating on two medical imaging datasets, the discovered circuits outperform commonly used encoding strategies while showing competitive results compared to purely classical counterparts.
We further analyze metrics to identify predictors of encoding performance. Entanglement capability and Fourier decomposition provide minimal insight, whereas the effective rank of the feature maps exhibits meaningful correlation and can serve as a threshold criterion to accelerate the search for high-performing encodings. 
Our findings provide both a practical method for encoding discovery and new insights into what makes data encodings effective in QML.
\end{abstract}


\fancypagestyle{specialfooter}{%
  \fancyhf{}
  \renewcommand\headrulewidth{0pt}
  \fancyfoot[R]{ \noindent\fbox{%
    \parbox{\textwidth}{%
        {\footnotesize This work has been submitted to the IEEE for possible publication. Copyright may be transferred without notice, after which this version may no longer be accessible.}
        }
    }}
}

\thispagestyle{specialfooter}

\input{sections/00_intro}

\input{sections/01_related_work}
\input{sections/02_background}
\input{sections/03_Encoding_search}

\input{sections/04_results}
\input{sections/05_conclusion}


\newpage
\printbibliography

\end{document}

%% file: sections/00_intro.tex
\section{Introduction}
Quantum-classical machine learning models stand as a promising way to leverage quantum computing to represent and perform computation in Hilbert space. This distinctive capacity has led researchers to investigate whether quantum models could offer different or more efficient representations of certain data structures compared to their classical counterparts~\cite{schuldEffectDataEncoding2021} or provide an intuition for quantum-inspired methods~\cite{tschöpe2025boostingclassificationquantuminspiredaugmentations}. The structure of these hybrid models can vary widely: the quantum component can be placed in front~\cite{matic2022quantum}, behind~\cite{monnet2024understanding}, or in between \cite{sakhnenko_hybrid_2022} classical neural networks. 

In this paper, we focus on quantum-classical convolutional neural networks (QCCNNs), an architecture studied for its reported performance~\cite{huang2023hybrid, matic2022quantum, monnet2024understanding}, generalization capabilities~\cite{6ckt-hlh8}, and near-term hardware readiness, as the shallow quantum circuits comprising the convolutional filters fit within the constraints of currently available devices. Hardware noise remains a practical concern for near-term deployment; in this work we operate in the noise-free simulation regime.

Beyond these practical considerations, a fundamental open question persists for hybrid quantum-classical models: to what extent does performance stem from the parameterization of the quantum circuit, and to what extent is it determined by how classical data is encoded into it? Disentangling these two contributions is non-trivial, since both the encoding and the variational parameters are typically optimized simultaneously. In this paper, we address this question by offloading  the optimization burden to a classical part searching over encoding strategies using a Monte Carlo tree search (MCTS) algorithm.

A natural question at this point is whether the effort of optimizing the encoding of a quantum-classical model is justified by any theoretical performance advantage over purely classical alternatives. Theoretical analyses suggest favorable generalization bounds for quantum-classical neural networks under certain conditions~\cite{6ckt-hlh8, caro_generalization_2022}, particularly in low-data regimes, but do not provide practical design guidance. Instead of approaching the problem from a quantum-theoretical perspective, we directly examine the encoded representations and assess feature space richness. The effective rank~\cite{roy2007effective} of the feature maps, quantifying diversity via the singular value spectrum, provides a data-driven metric computable without full model retraining. As shown in Section~\ref{sec: results}, this metric can guide encoding 
selection.

We propose the use of MCTS to navigate the space of possible data encoding schemes for QCCNNs. By systematically evaluating candidate encodings and correlating the task-relevant metrics computed for each with the resulting classification accuracy, we determine which metrics are predictive of use case performance and which are not. This finding has practical consequences for encoding design that extend beyond the specific architecture studied here.

Our contributions are thus threefold:

\begin{itemize}
    \item A MCTS algorithm for the automated discovery of data encoding strategies in QCCNNs, framing encoding selection as a sequential decision problem over gate composition.
    \item An empirical evaluation showing that MCTS-optimized encodings yield competitive performance against other quantum encodings and classical models on BreastMNIST and PneumoniaMNIST datasets, demonstrating that encoding strategy accounts for a significant share of the performance gap between our model and previously proposed strategies.
    \item A systematic analysis establishing the effective rank of feature maps as a predictive metric for encoding quality, providing a data-driven criterion for encoding selection that does not require 
    full model training.
\end{itemize}

%% file: sections/01_related_work.tex
\section{Related Work}\label{sec:related works}

\subsection{QCCNNs and Data Encoding}
Quantum convolutional neural networks (QCNNs)~\cite{cong2019quantum} adapt the convolutional structure of classical convolutional neural networks (CNNs) to the quantum domain, using quantum circuits to process data encoded in quantum states. However, current NISQ hardware limitations motivate hybrid quantum-classical approaches, where only specific components are executed on a quantum device. In such hybrid QCCNNs, a quantum circuit serves as the feature extractor and is combined with classical layers for classification. The quantum component can be either trainable, using parameterized quantum circuits, or fixed, using non-trainable circuit blocks. Recent works have applied QCCNNs to medical imaging data~\cite{xiangQuantumClassicalHybrid2024, simen2024digital, matic2022quantum}, motivating further investigation into their design. From a theoretical standpoint, as in classical machine learning, it remains difficult to favor one architecture over another based purely on theory. Substantial work has characterized the generalization capabilities of hybrid quantum models, including studies on effective dimension~\cite{abbas_power_2021} and generalization bounds~\cite{6ckt-hlh8, caro_generalization_2022}, but these do not prescribe specific design choices.

Training variational quantum circuits is computationally expensive and prone to barren plateaus~\cite{thanasilpSubtletiesTrainabilityQuantum2023, Cerezo_2022, Cerezo_2021}, motivating an encoding-only regime, where the quantum circuit serves as a fixed feature transformer and only the classical layers are trained~\cite{simen2024digital}. This sidesteps trainability issues and enables clearer analysis of encoding effects. While prior work shows that quantum advantage in kernel methods requires encoding feature maps that are hard to simulate classically~\cite{havlivcek2019supervised}, we do not bias the search towards such circuits; instead, our goal is to understand which encoding properties lead to good classification performance. This allows us to investigate whether hard-to-simulate encodings are necessary for strong performance, or whether efficient classical simulation methods could achieve comparable results.

The data encoding strategy significantly influences model performance~\cite{schuldEffectDataEncoding2021}. Despite its importance, the encoding properties that lead to good performance remain poorly understood. Commonly used metrics such as entanglement capability or expressibility show only weak correlations with predictive accuracy~\cite{monnet2024understanding, bowlesBetterClassicalSubtle2024}. Fourier-based analyses suggest that spectral properties may provide more informative indicators~\cite{stroblFourierFingerprintsAnsatzes2025, wiedmannFourierAnalysisVariational2024, schuldEffectDataEncoding2021}.

\subsection{Quantum Circuit Design}
Automated quantum circuit design has been explored across diverse application domains, including quantum chemistry~\cite{mengQuantumCircuitArchitecture2021a, ryabinkinIterativeQubitCoupled2019}, combinatorial optimization~\cite{foderaReinforcementLearningVariational2024, turatiAutomatedDesignStructured2025}, and QML~\cite{nguyenQuantumEmbeddingSearch2022, rappReinforcementLearningbasedArchitecture2025, altares-lopezAutomaticDesignQuantum2021, wuQuantumDARTSDifferentiableQuantum2023, luMarkovianQuantumNeuroevolution2021a}. Various search methodologies have been employed, including reinforcement learning~\cite{foderaReinforcementLearningVariational2024, rappReinforcementLearningbasedArchitecture2025, patelCurriculumReinforcementLearning2024}, evolutionary algorithms~\cite{altares-lopezAutomaticDesignQuantum2021, luMarkovianQuantumNeuroevolution2021a}, and Bayesian optimization~\cite{duongQuantumNeuralArchitecture2022, pirhooshyaranQuantumCircuitDesign2021, sunQuantumArchitectureSearch2024}. MCTS has emerged as a particularly effective approach for circuit design~\cite{wangAutomatedQuantumCircuit2023, mengQuantumCircuitArchitecture2021a}. Ref.~\cite{lipardiQuantumCircuitDesign2025} introduces progressive widening to MCTS for variational quantum circuit design, enabling efficient exploration of large action spaces while reducing the number of circuit evaluations by up to two orders of magnitude compared to prior methods.

In the context of QML, existing work predominantly focuses on searching for variational ansatz architectures or entangling layers, while the data encoding strategy is typically chosen from already existing schemes rather than optimized~\cite{nguyenQuantumEmbeddingSearch2022, wuQuantumDARTSDifferentiableQuantum2023, luMarkovianQuantumNeuroevolution2021a}. To the best of our knowledge, this is the first study to focus exclusively on automated data encoding circuit design for QCCNNs, entirely removing the variational component. Using MCTS with progressive widening, we discover fixed encoding circuits that serve as feature extractors for classical classifiers, enabling a systematic analysis of which encoding properties lead to improved classification performance.

%% file: sections/02_background.tex
\section{Background} \label{sec:background}
\subsection{Datasets}
We focus on medical imaging data, which characteristically comprise relatively small datasets, typically ranging from a few hundred to a few thousand samples. These are precisely the data regimes in which the generalization capabilities of quantum models are expected to be most relevant~\cite{caro_generalization_2022}. We select two binary classification datasets from the MedMNIST collection~\cite{medmnistv1, medmnistv2}:

BreastMNIST contains 546 training, 78 validation, and 156 testing 2D breast ultrasound images categorized as malignant or non-malignant, where the non-malignant class is obtained by merging normal and benign cases.

PneumoniaMNIST contains 4708 training, 524 validation, and 624 testing 2D chest X-ray images categorized as pneumonia or normal. To reduce encoding time, we randomly sample a subset of 800 images while maintaining the original class distribution.

For both datasets, pixel values are normalized to the range $[-1, 1]$. We primarily use the downscaled 28$\times$28 images to minimize encoding time. To investigate whether the derived encoding circuits retain good performance at higher resolutions, we also consider the 128$\times$128 images.

\subsection{Classical Models}

The classical baselines consist of a fully connected (FC) network and a simple CNN. The FC model flattens the input image and passes it through a single fully connected layer with 1570 trainable parameters. The CNN adds one convolutional layer with $k \times k = 2 \times 2$ filter size, $c = 1$ input channel, and $n = 4$ output filters, followed by ReLU activation and a fully connected classification layer, adding $(k^2 \times c + 1) \times n = 20$ parameters to the FC baseline. This CNN architecture is chosen to match the QCCNN structure as closely as possible, using the same filter size and number of output feature maps, enabling a fair comparison between classical convolutional and quantum encoding-based feature extraction.

Both models are implemented in PyTorch~\cite{paszke2019pytorchimperativestylehighperformance}. Training uses the Adam optimizer with a learning rate of $5 \times 10^{-4}$ and batch size of 32.

\subsection{Performance metrics}
Identifying properties of data encoding circuits that correlate with classification performance could enable more efficient circuit design by filtering out poor candidates before expensive evaluation. We investigate three metrics: entanglement capability, fourier spectrum, and feature map diversity. 
 
\subsubsection{Entanglement capability}

The entanglement capability~\cite{Sim_2019} quantifies the level of entanglement a circuit can produce, ranging from 0 (product states with no entanglement) to 1 (highly entangled states). It is based on the Meyer-Wallach entanglement measure $Q(|\psi\rangle)$, averaged over a sampled set $S$ of parameter vectors:
\begin{equation}
    \text{Ent} = \frac{1}{|S|} \sum_{\theta_i \in S} Q(|\psi_{\theta_i}\rangle),
\end{equation}
where in our case the parameters $\theta_i$ correspond to the encoded data inputs.

\subsubsection{Fourier spectrum}
Quantum circuits with data encoding gates can be understood through a Fourier series representation~\cite{schuldEffectDataEncoding2021}. For encoding gates of the form $S(x) = e^{ixH}$, where $x \in \mathbb{R}$ is the input and $H$ is a Hamiltonian, the function describing the quantum model can be written as
\begin{equation}
    f(x) = \sum_{\omega \in \Omega} c_\omega e^{i\omega x},
\end{equation}
where $\omega$ is a frequency in the spectrum $\Omega$ determined by the eigenvalues of the encoding Hamiltonians, and the coefficients $c_\omega \in \mathbb{C}$ satisfy $c_\omega = c^*_{-\omega}$.

Several studies have shown that spectral properties may provide informative indicators of performance in variational settings~\cite{monnet2024understanding, stroblFourierFingerprintsAnsatzes2025, wiedmannFourierAnalysisVariational2024}. However, in our setting, the coefficients are entirely determined by the circuit architecture and the input data, rather than being adjustable through training. This raises the question of whether Fourier spectral properties remain predictive when coefficients cannot be optimized.

\subsubsection{Feature Map Diversity} \label{sec:effective_rank}
In classical CNNs, feature diversity is known to improve generalization~\cite{cogswellReducingOverfittingDeep2016}, and redundant information contributes little to model performance~\cite{liPruningFiltersEfficient2017}. Rank-based metrics have been used to identify redundant filters in CNNs~\cite{lin2020hrank}. We use the effective rank~\cite{roy2007effective}, an entropy-based measure of matrix diversity, of the feature maps produced by quantum encoding circuits to gauge how feature diversity influences performance.

For a matrix $A$ with singular values $\sigma_1, \sigma_2, \dots, \sigma_k$, 
the effective rank is defined as
\begin{equation}
    \text{erank}(A) = \exp\left( -\sum_{i=1}^{k} p_i \log p_i \right),
\end{equation}
where $p_i = \sigma_i / \sum_{j=1}^{k} \sigma_j$ represents the singular value distribution. This formulation corresponds to the exponential of the Shannon entropy of the singular value spectrum. A matrix with a single dominant singular value has low effective rank, indicating redundancy, while a matrix with uniformly distributed singular values has high effective rank, indicating diverse and independent components.

To compare circuits producing different numbers of feature maps, we normalize the effective rank to the interval $[0, 1]$:
\begin{equation}
    \text{erank}_{\text{norm}}(A) = \frac{\text{erank}(A) - 1}{m - 1},
\end{equation}
where $m$ is the number of feature maps. A value of 1 indicates that all feature channels contain independent information, whereas a value approaching 0 indicates high redundancy, where multiple channels could be collapsed without significant information loss.

%% file: sections/03_Encoding_search.tex
\section{Data encoding search via MCTS}\label{sec:method}
\subsection{QCCNN Architecture}
Typical QCCNNs consist of a data encoding circuit, a trainable variational circuit, and measurements, possibly fed into a classical network~\cite{monnet2024understanding}. In contrast, our architecture removes the variational circuit entirely: the quantum component acts as a fixed feature extractor using automatically designed data encoding circuits, while all trainable parameters reside in the classical layers. 

\begin{figure}[!htbp]

    \includegraphics[width=0.51\textwidth]{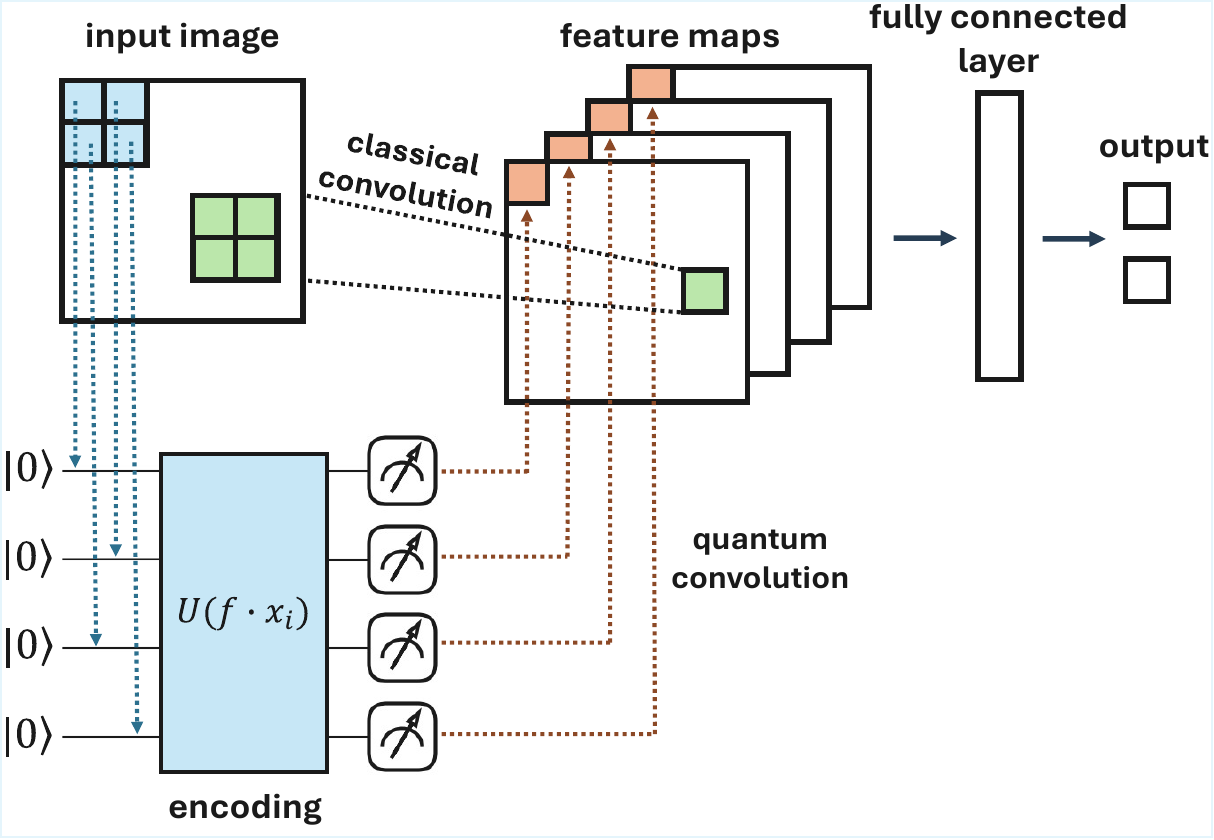}
    
    \caption{Architecture sketch of the CNN and QCCNN for the patch size $2\times2$.}
    \label{fig:QCCNN}
\end{figure}

Analogous to classical CNNs, a $k \times k$ filter slides over the image with stride $k$. Each patch of $k^2$ pixels is encoded into $k^2$ qubits using rotation gates $R_i(f \cdot x)$ with $i \in \{X, Y, Z\}$, where angles are determined by pixel values $x$ scaled by a factor $f$, and two-qubit gates $R_{ZZ}(f \cdot x_i \cdot x_j)$ encoding pairwise pixel interactions. The scaling factor $f$ adjusts the effective range of rotation angles and can significantly influence model performance~\cite{monnet2024understanding}. Hadamard and CNOT gates serve as auxiliary operations. We consider filter sizes of $2 \times 2$ (4 qubits) and $3 \times 3$ (9 qubits) with strides 2 and 3. After encoding, each qubit is measured in the $Z$-basis, yielding $k^2$ feature maps that are flattened and passed to a fully connected layer for classification, as seen in \Cref{fig:QCCNN}. The quantum encoding circuits remain fixed after discovery by the search procedure, only the classical layer is trained.

The quantum circuits are simulated using PennyLane~\cite{bergholm2022pennylaneautomaticdifferentiationhybrid} in a noiseless environment. The same training configuration as the classical models is used for the classical layers of the QCCNN.

\subsection{Data encoding search}
Designing effective data encoding circuits for quantum machine learning is challenging due to the vast search space of possible circuit architectures. Manual design requires expert knowledge and extensive trial-and-error, while exhaustive search is computationally infeasible. Inspired by recent work on automated quantum circuit design~\cite{lipardiQuantumCircuitDesign2025}, we employ MCTS with progressive widening to automatically discover high-performing data encodings for our QCCNN.

We formulate the data encoding search as a sequential decision problem. Each node in the search tree represents a candidate encoding circuit, and edges correspond to circuit modifications. The gate pool consists of
\begin{equation}
    \mathcal{G} = \{ R_X, R_Y, R_Z, R_{ZZ}, H, \text{CNOT} \},
\end{equation}
where the parameterized gates encode the input pixel values of the image patch. At each step, the search can perform one of three actions:
\begin{enumerate}
    \item \textbf{Add:} Append a gate from $\mathcal{G}$ to the circuit.
    \item \textbf{Remove:} Delete an existing gate from the circuit.
    \item \textbf{Replace:} Substitute an existing gate with a different 
    gate from $\mathcal{G}$.
\end{enumerate}
Actions are sampled according to probabilities that depend on the current circuit size. During the initial phase, when a circuit contains fewer than $2n$ gates (where $n$ is the number of qubits), only add actions are performed to ensure circuits reach a minimum complexity. Once this threshold is reached, actions are sampled with probabilities: add (40\%), remove (15\%), and replace (45\%). When the circuit reaches the maximum allowed number of gates, add actions are disabled and only remove (50\%) and replace (50\%) actions are sampled. These probabilities were chosen heuristically to balance exploration of new structures with refinement of existing circuits.

Our MCTS implementation follows the UCT (Upper Confidence bounds applied to Trees) four-phase structure~\cite{kocsisBanditBasedMonteCarlo2006}:
\begin{enumerate}
    \item \textbf{Selection:} Starting from the root node, child nodes are selected by maximizing the Upper Confidence Bound (UCB1)~\cite{auerFinitetimeAnalysisMultiarmed2002}:
    \begin{equation}
        \text{UCB1}(s, a) = \bar{Q}(s, a) + c \sqrt{\frac{\ln N(s)}{N(s, a)}},
    \end{equation}
    where $\bar{Q}(s, a)$ is the cumulative reward for taking action $a$ in state $s$, $N(s)$ is the visit count of state $s$, $N(s, a)$ is the number of times action $a$ was taken from $s$, and $c$ is the exploration constant balancing exploitation and exploration.
    
    \item \textbf{Expansion:} When a leaf node is reached, new child nodes are added by sampling actions from the action space.
    
    \item \textbf{Simulation:} The candidate circuit is evaluated by training the QCCNN and computing the validation area under the ROC curve (AUC).
    
    \item \textbf{Backpropagation:} The reward is propagated back through the tree, updating visit counts and average rewards along the path.
\end{enumerate}

Following~\cite{lipardiQuantumCircuitDesign2025}, we employ progressive widening to manage the large action space. Rather than enumerating all possible actions at each node, new actions are added progressively as the visit count of the node increases: 
\begin{equation}
    |A(s)| \leq k \cdot N(s)^\alpha,
\end{equation}
where $k$ and $\alpha$ are hyperparameters controlling the widening rate. This allows the search to focus computational resources on promising regions of the search space while gradually exploring alternatives.

The reward for a candidate encoding circuit is defined as the validation AUC achieved by the QCCNN after training. Since both datasets are imbalanced, AUC provides a more reliable measure of classification performance than accuracy. Higher AUC corresponds to higher reward, guiding the search toward encodings that yield better generalization.

The search is initialized with a single layer of $R_X$ gates on all qubits. Each data encoding candidate is evaluated by training the 
QCCNN for 30 epochs with a single weight initialization, using the best validation AUC as the reward value. The exploration constant is set to $c = 0.4$, and the progressive widening parameters are $k = 1$ 
and $\alpha = 0.3$.

%% file: sections/04_results.tex
\section{Results} \label{sec: results}
We present the results of the data encoding search for our QCCNN model on two medical imaging datasets. We then compare its performance with two classical models: one using only a fully connected layer and one using a CNN. As performance metrics, we focus on the cross-entropy loss, accuracy, and AUC for the training, validation, and test sets. The performance of the different data encoding circuits is analyzed in terms of various metrics, to identify which properties are indicative of better performance. 

\subsection{Performance of the found encoding circuits} \label{sec:performance_encodings}
\begin{figure*}[!htbp]
    \centering
    \begin{subfigure}[b]{0.45\textwidth}
        \centering
        \includegraphics[width=\textwidth]{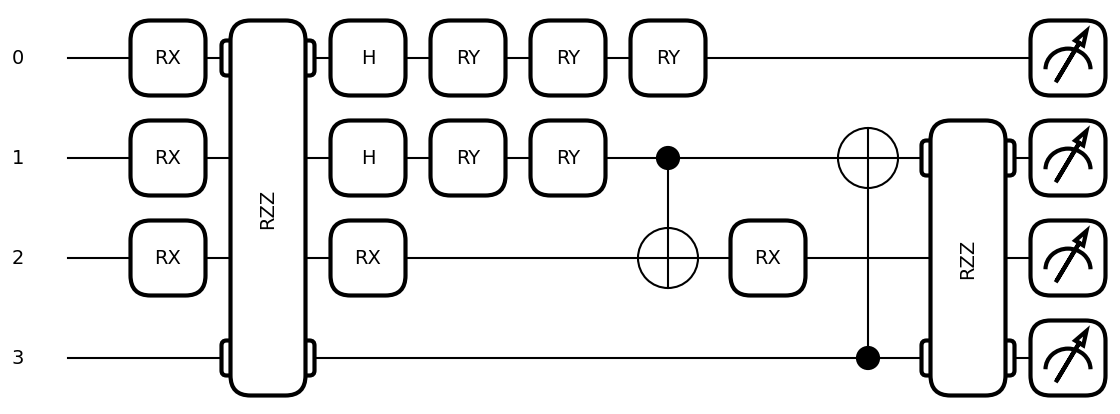}
        \caption{BreastMNIST 2$\times$2}

    \end{subfigure}
    \hspace{0.5cm}  
    \begin{subfigure}[b]{0.43\textwidth}
        \centering
        \includegraphics[width=\textwidth]{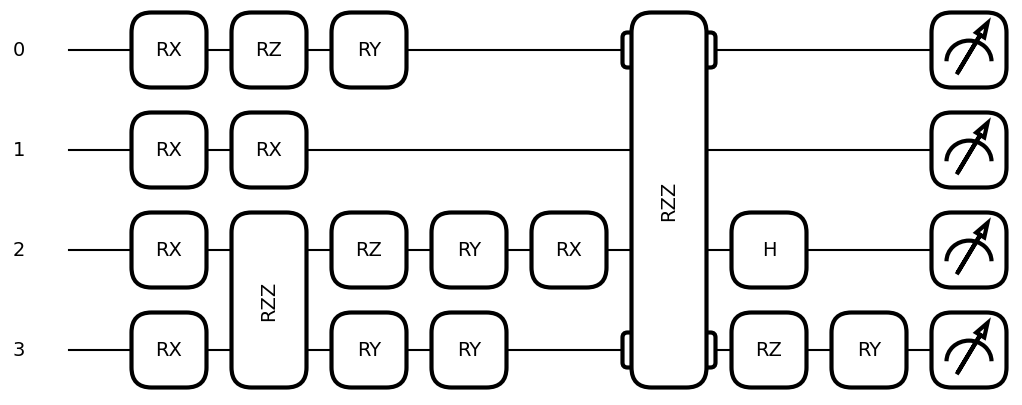}
        \caption{PneumoniaMNIST 2$\times$2}
    \end{subfigure}
    
    \vspace{0.3cm}
    
    \begin{subfigure}[b]{0.43\textwidth}
        \centering
        \includegraphics[width=\textwidth]{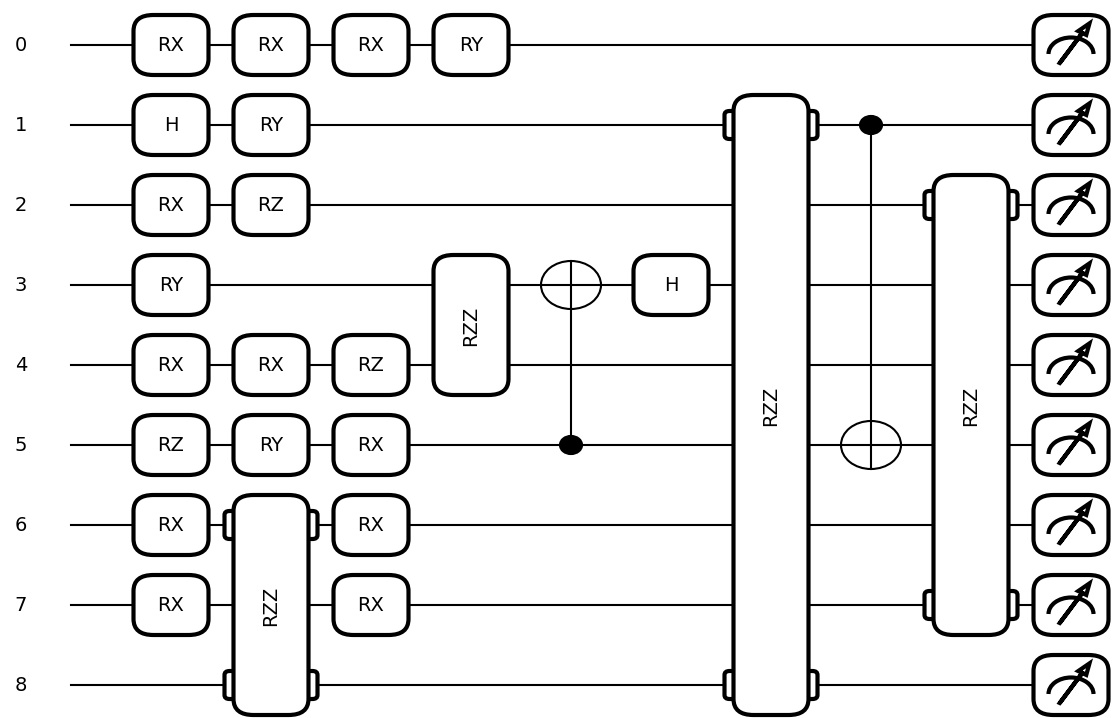}
        \caption{BreastMNIST 3$\times$3}
    \end{subfigure}
    \hspace{0.5cm}  
    \begin{subfigure}[b]{0.35\textwidth}
        \centering
        \includegraphics[width=\textwidth]{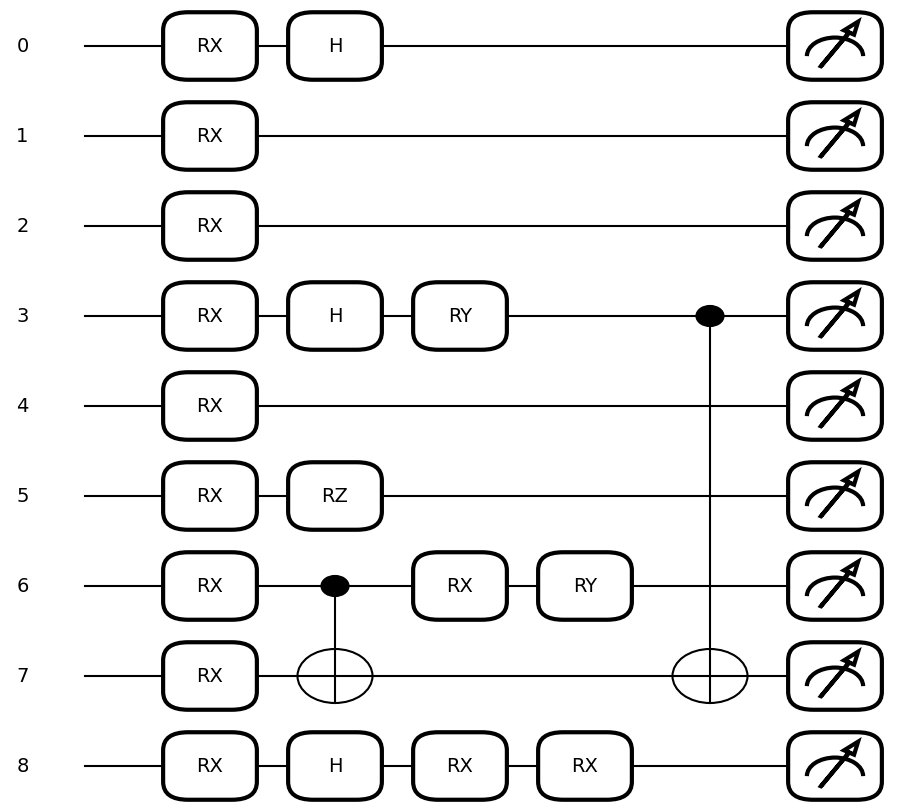}
        \caption{PneumoniaMNIST 3$\times$3}

    \end{subfigure}
    
    \caption{Data encoding circuits with the highest validation AUC identified through MCTS for BreastMNIST (a, c) and PneumoniaMNIST (b, d) using 2$\times$2 (top) and 3$\times$3 (bottom) patch sizes. }
    \label{fig:data_encodings}
\end{figure*}

We search for optimal data encodings for both datasets using 2$\times$2 and 3$\times$3 patch sizes. After the search, we select the best-performing encodings based on validation AUC. The resulting circuits for each dataset and patch size are shown in \Cref{fig:data_encodings}.

After the search, we optimize the scaling factor $f$ of the input $x$ for the best-found data encoding circuits. As shown in~\cite{monnet2024understanding}, this step can significantly influence the resulting performance. The scaling factor is selected from the interval $f \in [0.5, 2.0]$ based on validation performance and used for all subsequent analysis. For the BreastMNIST dataset, we use $f_{2\times2} = 1.0$ and $f_{3\times3} = 0.8$, and for the PneumoniaMNIST dataset, we use $f_{2\times2} = 0.9$ and $f_{3\times3} = 1.1$.

\begin{figure*}[!htbp]
    \centering
    \includegraphics[width=0.9\textwidth]{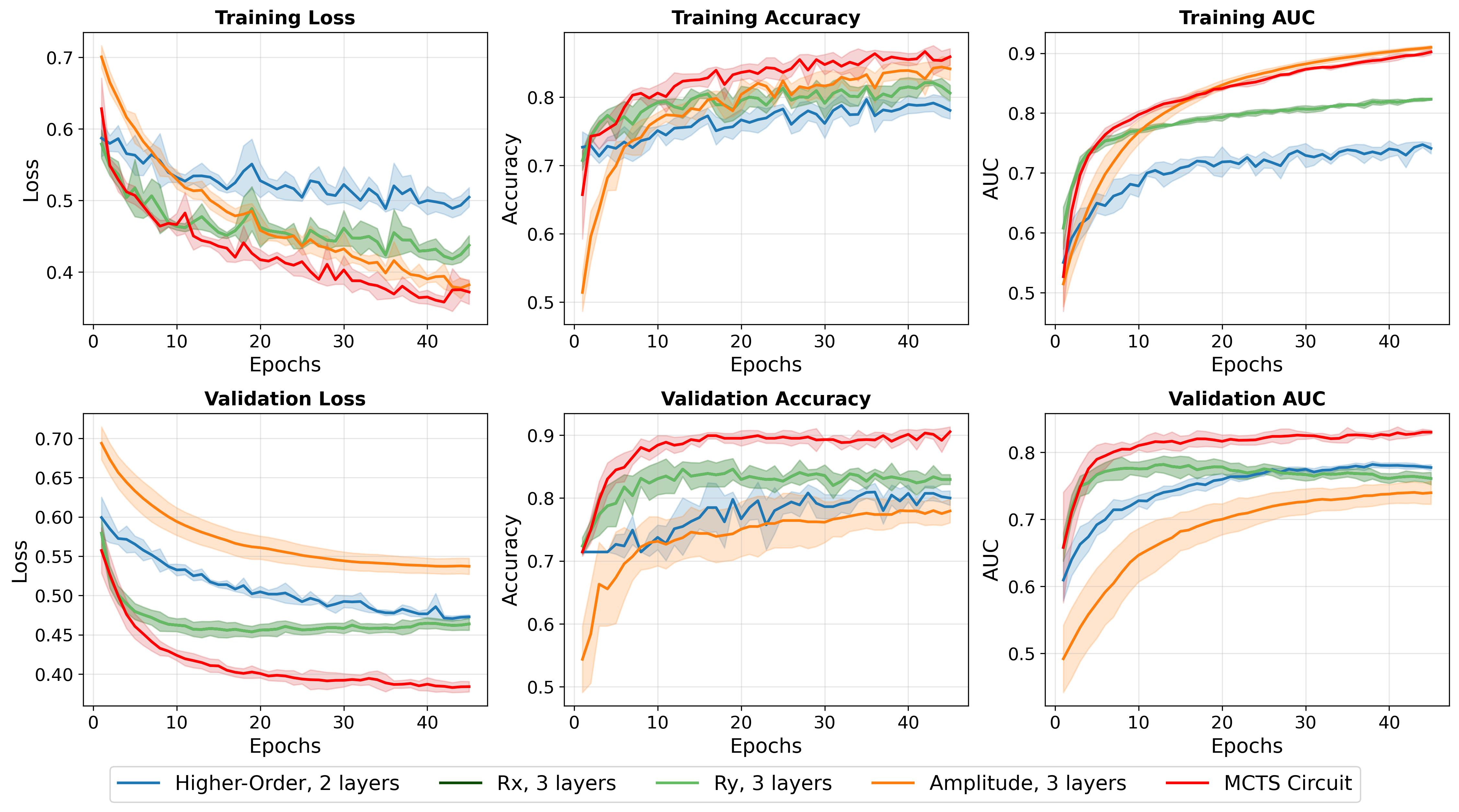}
    
    \caption{Comparison of data encoding strategies on BreastMNIST. The MCTS-derived 2$\times$2 circuit is compared against baseline encodings using $R_X$, $R_Y$, and higher-order gates with 1 to 3 re-uploading layers; only the best-performing configuration for each baseline is shown. The two angle encodings ($R_X$ and $R_Y$) achieve identical performance and therefore overlap. Results are averaged over five random seeds, with error bars indicating standard deviation.}
    \label{fig:encoding_comparison_breast}
\end{figure*}

Next, we assess whether the discovered circuits outperform existing data encoding strategies. We consider angle encoding~\cite{weigold2021encoding} using $R_X$ and $R_Y$ gates, as well as higher-order encoding~\cite{havlivcek2019supervised} and amplitude encoding~\cite{weigold2021encoding}. For each approach, we vary the number of data re-uploading layers from one to three and select the best-performing configuration for comparison. \Cref{fig:encoding_comparison_breast} compares the baseline encodings to the 2$\times$2 circuit identified by the MCTS for the BreastMNIST dataset. The MCTS-derived circuit significantly outperforms all baseline methods across all metrics on both the training and validation sets. This demonstrates that the choice of data encoding strongly influences model performance and that automated circuit search is a valuable component of the overall approach.

\begin{figure*}[!htbp]
    \centering
    \begin{subfigure}[b]{\textwidth}
        \centering
        \includegraphics[width=0.95\textwidth]{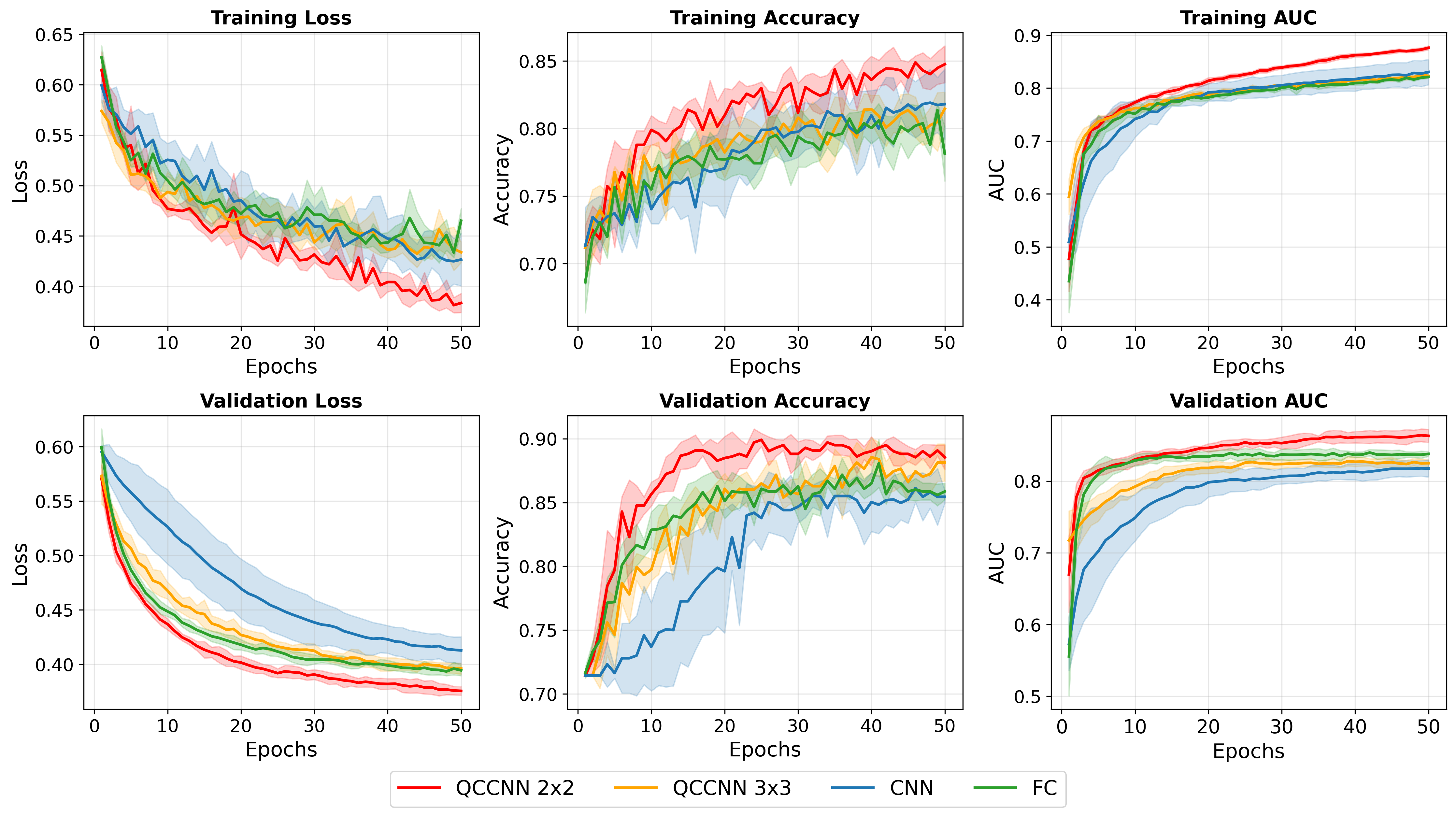}
        \caption{BreastMNIST}
        \label{fig:model_comp_breast}
    \end{subfigure}
    
    \vspace{0.4cm}
    
    \begin{subfigure}[b]{\textwidth}
        \centering
        \includegraphics[width=0.95\textwidth]{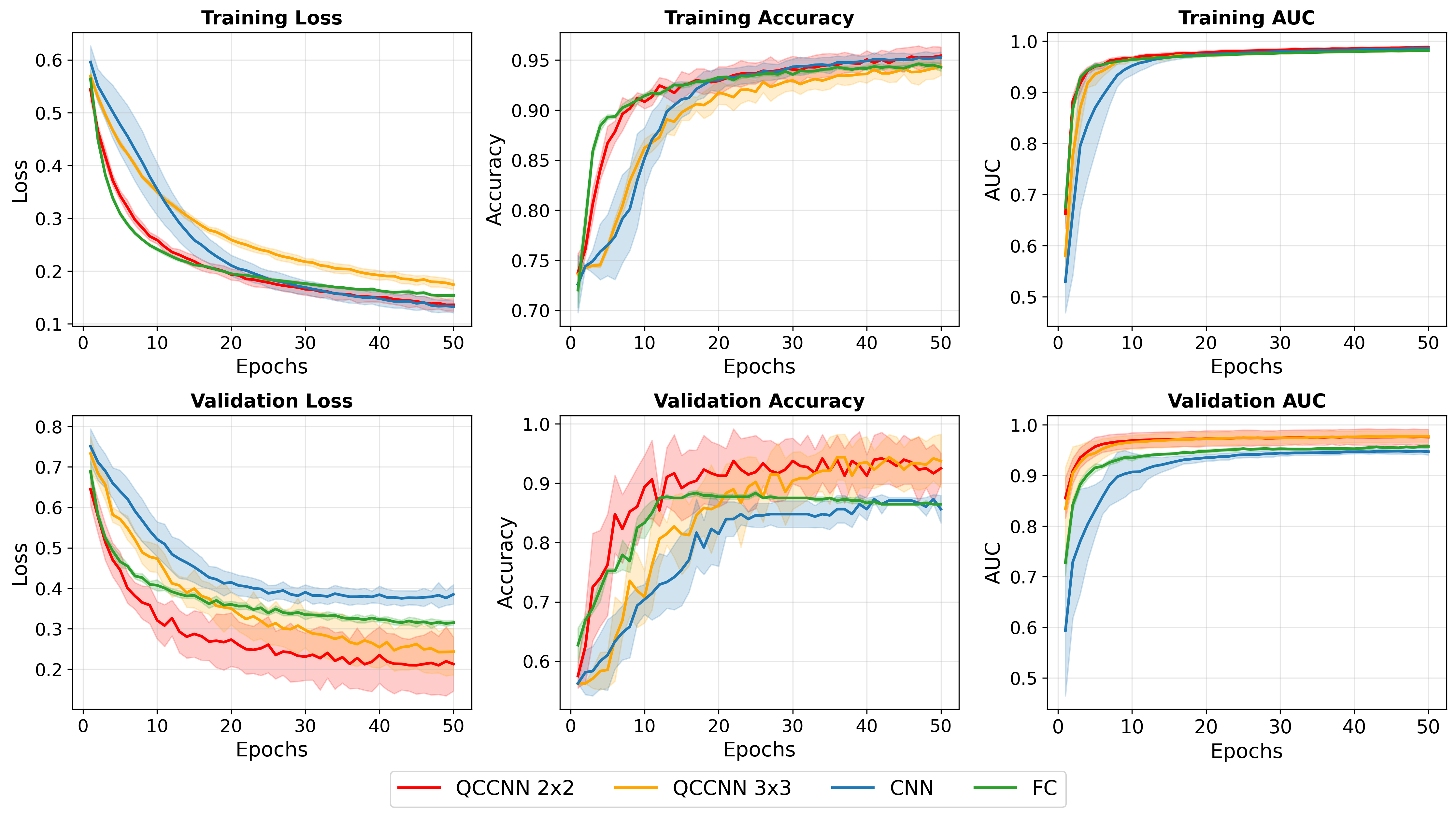}
        \caption{PneumoniaMNIST}
        \label{fig:enc_comp_breast}
    \end{subfigure}
    
    \caption{Performance evaluation on (a) BreastMNIST  and (b) PneumoniaMNIST datasets. Training metrics (top row of each subplot) and validation metrics (bottom row) shown as mean $\pm$ standard deviation over 5 seeds.}
    \label{fig:model_comparison}
\end{figure*}

We next compare the performance of the best-found data encodings for both 2$\times$2 and 3$\times$3 patch sizes against the classical models. Each model is trained five times with different weight initializations; reported metrics are averaged over seeds with error bands indicating standard deviation. \Cref{fig:model_comparison} shows the loss, accuracy, and AUC for the QCCNN variants and the classical models on both the BreastMNIST and PneumoniaMNIST datasets.

Across both datasets, the QCCNN with 2$\times$2 patches consistently achieves the best or near-best validation performance across all metrics, while the CNN performs worst and converges more slowly (\Cref{fig:model_comparison}). The 3$\times$3 QCCNN shows greater variability across seeds, particularly on PneumoniaMNIST where some initializations reach near-perfect accuracy. The performance gap between QCCNN and classical models differs by dataset. On PneumoniaMNIST, both QCCNN variants substantially outperform the classical baselines across all metrics. On BreastMNIST, the advantage is more modest: the 3$\times$3 QCCNN performs comparably to the FC model, which achieves higher AUC than the larger patch configuration.

Overall, the automatically discovered encoding circuits enable the QCCNN to match or exceed classical baselines despite comparable parameter counts, suggesting that quantum feature extraction provides meaningful representations for these medical imaging tasks.

\begin{figure}[!htbp]
  \centering
  \begin{subfigure}[b]{0.7\columnwidth}
       \centerline{\includegraphics[width=0.93\columnwidth]{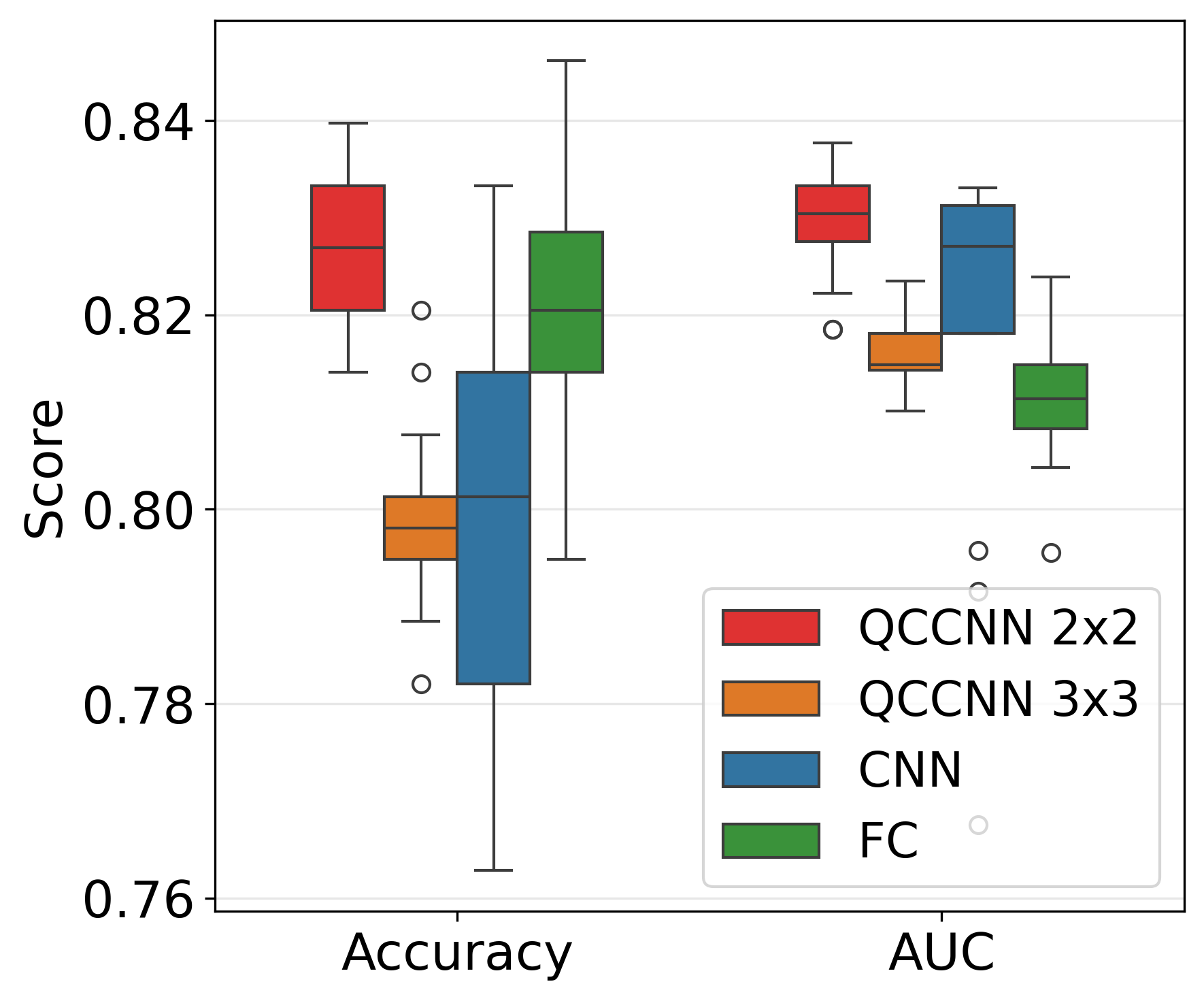}}
      \caption{BreastMNIST}
      \label{fig:test_breast}
  \end{subfigure}
  
  \vspace{0.15cm}

  \centering
  \begin{subfigure}[b]{0.7\columnwidth}
       \centerline{\includegraphics[width=0.93\columnwidth]{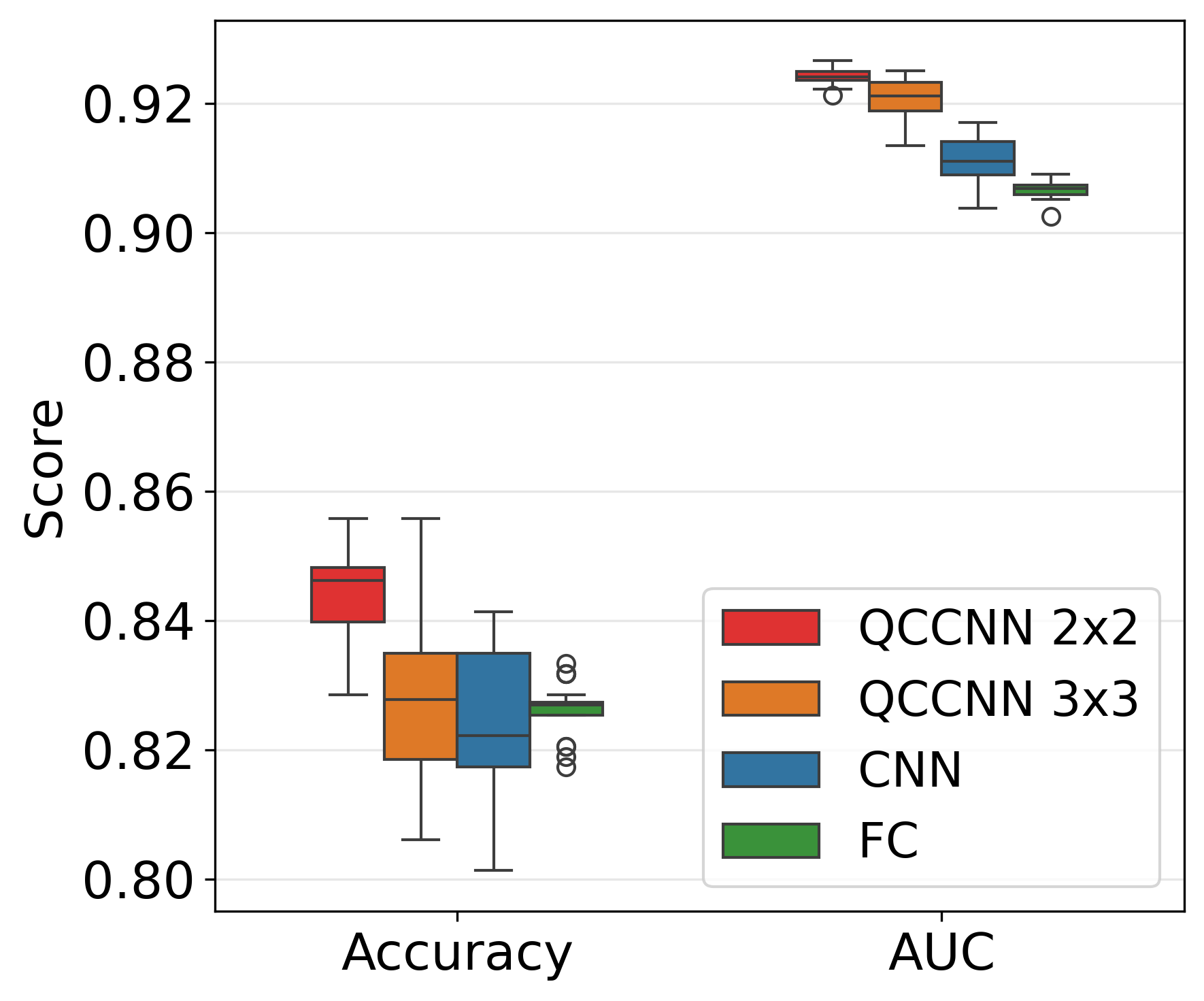}}
      \caption{PneumoniaMNIST}
      \label{fig:test_pneu}
  \end{subfigure}
    
    \caption{Accuracy and AUC performance on the (a) BreastMNIST and (b) PneumoniaMNIST test datasets. Each model is evaluated over 20 random weight initializations to capture variability across different seeds.}
    \label{fig:test}
\end{figure}

To evaluate the performance of the various models on unseen data, we focus on the test datasets. \Cref{fig:test} reports the resulting statistical distribution of accuracy and AUC values, obtained from 20 random weight initializations for each model. In the following, we distinguish between median performance across initializations and the best performance achieved for a specific seed.

Across both datasets, the QCCNN with 2$\times$2 patches achieves the highest median accuracy and AUC, while also exhibiting the narrowest spread between minimum and maximum values, indicating both strong typical performance and robustness to initialization. The CNN shows highest sensitivity to initialization with the widest performance range. The FC model performs comparably to the CNN in median metrics but with moderately better stability. The datasets differ primarily in best-seed results: for BreastMNIST, the FC model achieves the highest single-seed accuracy while QCCNN 2$\times$2 reaches the highest AUC; for PneumoniaMNIST, both QCCNN variants achieve the highest accuracy, with the 2$\times$2 configuration also attaining the highest AUC. The 3$\times$3 QCCNN performs comparably to the CNN in median accuracy but with notably lower variability.

The consistently strong AUC performance of the QCCNN across both datasets suggests that quantum data encoding may help highlight relevant features in the images.

The images considered so far have a resolution of 28$\times$28. This raises the question: do the identified data encodings also improve performance at higher resolutions? To investigate this, we extend our analysis to 128$\times$128 images using the 2$\times$2 MCTS-derived encodings for each dataset. We select circuits at ranks 1, 400, and 700 by validation AUC (denoted $C_1$, $C_{400}$, and $C_{700}$), representing best, medium, and worst performing encodings. The quantum encoding circuits remain fixed from the 28$\times$28 search; only the fully connected layers are retrained.

\begin{figure}[!htbp]
    \begin{subfigure}[]{\columnwidth}
         \centerline{\includegraphics[width=0.95\columnwidth]{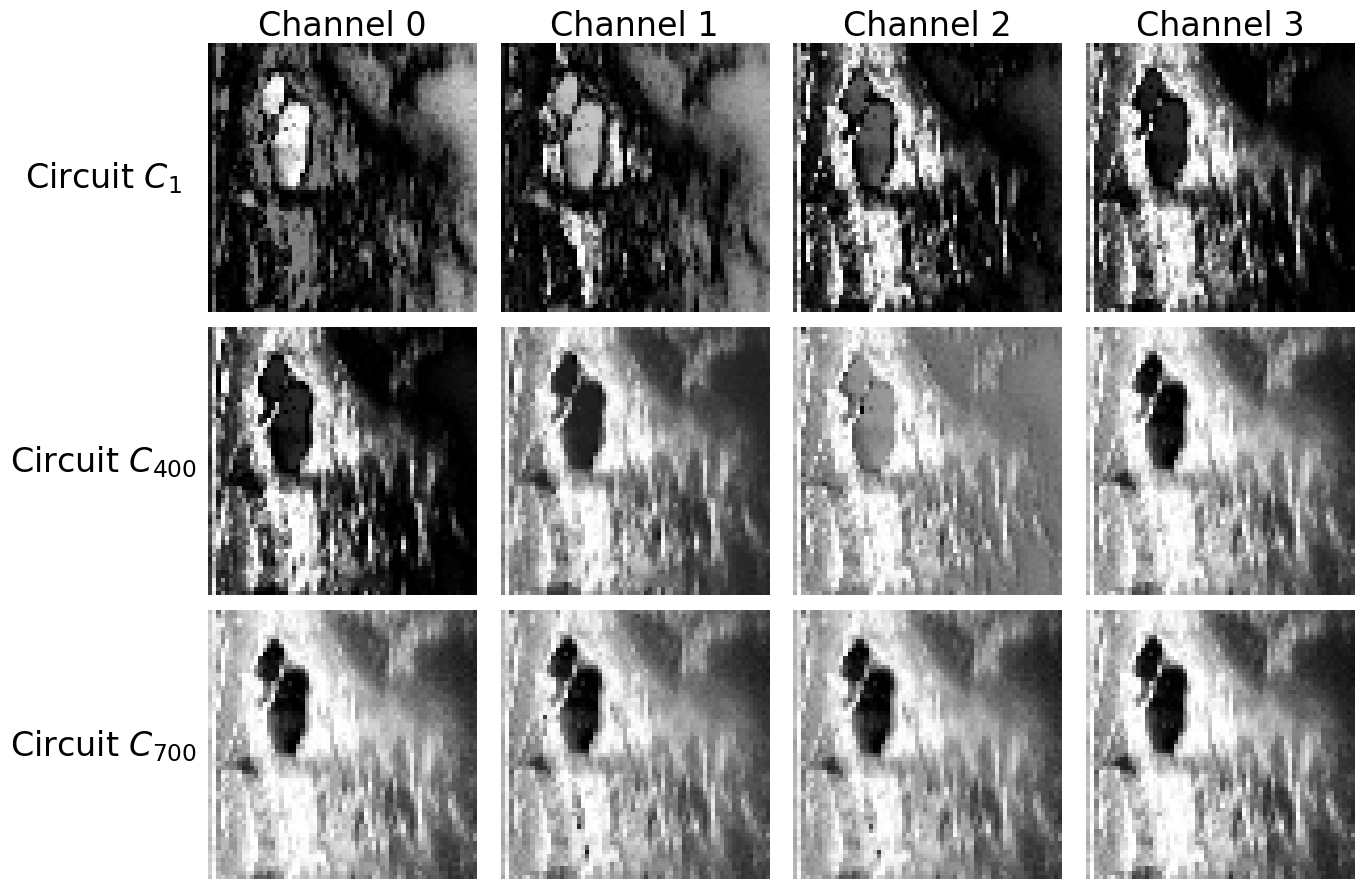}}
    \end{subfigure}
    \caption{Feature maps produced by 2$\times$2 patch-sized quantum 
    encodings on BreastMNIST. Circuits $C_1$, $C_{400}$, and $C_{700}$ 
    represent the best, medium, and worst performing encodings from 
    the search, respectively. Higher-ranked circuits produce more 
    diverse feature representations.}
    \label{fig:Feature_maps}
\end{figure}

The results for BreastMNIST, summarized in \Cref{tab:128x128}, confirm that the relative rankings among encodings are preserved: $C_1$ performs best, followed by $C_{400}$ and then $C_{700}$, across all three metrics. We observe similar behavior for PneumoniaMNIST. This suggests that encoding quality generalizes across image resolutions. To gain insight into why certain encodings outperform others, we visualize the feature maps after encoding and before the fully connected layer (\Cref{fig:Feature_maps}). The better-performing circuits produce more diverse feature representations, functioning similarly to classical convolutional layers by highlighting distinct structures within the images. This diversity likely aids classification by providing the model with richer, more discriminative information.

However, while encoding quality rankings transfer, the encodings optimized for 28$\times$28 images do not retain their advantage over classical models at higher resolutions. As shown in \Cref{tab:128x128}, the CNN achieves the best performance across all metrics on BreastMNIST, and we observe similar results for PneumoniaMNIST. Note that the CNN retrains both its convolutional filters and fully connected layer, giving it additional flexibility to adapt to the higher resolution, whereas the QCCNN uses fixed encoding circuits discovered at the lower resolution. This suggests that a dedicated encoding search at the target resolution may be required to realize potential quantum advantages at larger scales.

\begin{table}[!htbp]
    \centering
    \caption{Performance comparison of selected encoding circuits and 
    classical baselines on the 128$\times$128 BreastMNIST validation dataset. Results are averaged 
    over five seeds, values in parentheses indicate standard deviation. Best results are highlighted in bold.}
    \label{tab:128x128}
    \begin{tabular}{lccc}
        \toprule
        Model & Loss & Accuracy  & AUC  \\
        \midrule
        QCCNN $C_1$     & 0.389(16) & 0.880(7)  & 0.847(8)  \\
        QCCNN $C_{400}$ & 0.459(29) & 0.845(16) & 0.796(14) \\
        QCCNN $C_{700}$ & 0.463(19) & 0.843(18) & 0.792(4)  \\
        \midrule
        FC        & 0.362(12) & 0.872(9)  & 0.875(6)  \\
        CNN       & \textbf{0.350(41)} & \textbf{0.884(19)} & \textbf{0.883(23)} \\
        \bottomrule
    \end{tabular}
\end{table}

\subsection{Metric–Performance Correlation Analysis}
The following section investigates whether the performance of data encodings can be linked to specific metrics. Identifying such correlations could simplify the search for optimized encoding strategies by filtering out poor candidates before running computationally expensive performance evaluations. We systematically evaluate three different metrics: entangling capability, Fourier coefficients and feature map diversity.

\subsubsection{Entanglement Capability}
\begin{figure}[b]
        \centering
    \begin{subfigure}[a]{\columnwidth}
         \centerline{\includegraphics[width=0.75\columnwidth]{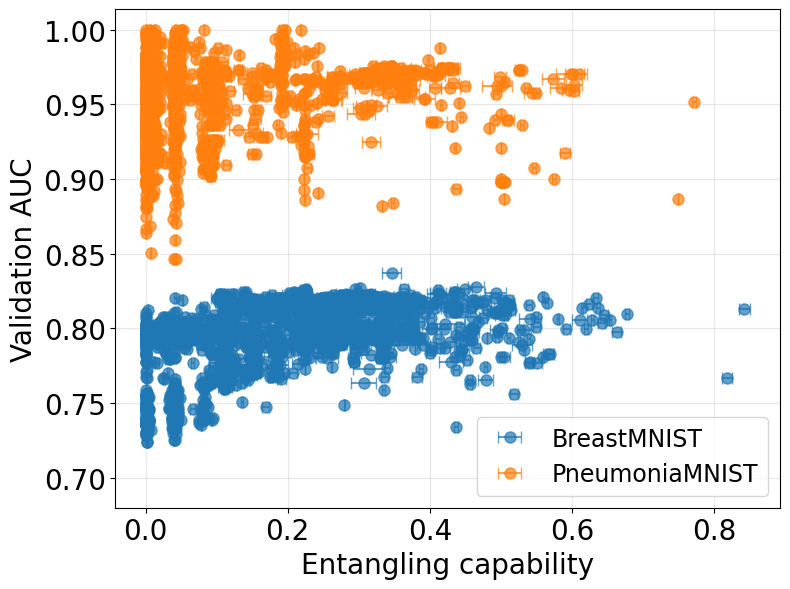}}
    \end{subfigure}

    \caption{Entanglement capability versus validation AUC of various encoding circuits for the BreastMNIST (blue) and the PneumoniaMNIST (orange) datasets. Entanglement values are averaged over three input parameters, with error bars indicating standard deviation.  }
    \label{fig:entanglement_capability}
\end{figure}

First, we examine the correlation between the entanglement capability and the performance of the various encodings. \Cref{fig:entanglement_capability} shows the entanglement capability against the validation AUC for encodings generated across multiple MCTS runs. Notably, the values mostly cluster around lower entanglement capabilities, indicating that the search is not driven towards highly entangling circuits.  The correlation between entanglement and performance is inconsistent across datasets. For BreastMNIST, we observe a moderate positive correlation 
($r = 0.49$, $p < 0.001$), while for PneumoniaMNIST only a weak correlation is found ($r = 0.08$), which, although statistically significant ($p = 0.003$), is not practically meaningful. Therefore, entanglement capability is not a reliable predictor of encoding quality across datasets. 

\subsubsection{Fourier coefficients analysis}
To extract the Fourier spectrum empirically, we use the discrete Fourier transform. For each qubit $q$ we fix the inputs to all other qubits at random values sampled uniformly from $[-1, 1]$ and sweep the input $x_q$ over a uniform grid of $N = 50$ points in the same interval. The circuit output for each qubit is decomposed with the real fast Fourier transform (RFFT) leading to the coefficients. To capture the effect of qubit interactions we repeat this 100 times for random input values and analyze the distribution of the first five coefficients, plus the zero frequency, on the complex plane.

\begin{figure}[b]
    \centering
    \includegraphics[width=1\columnwidth]{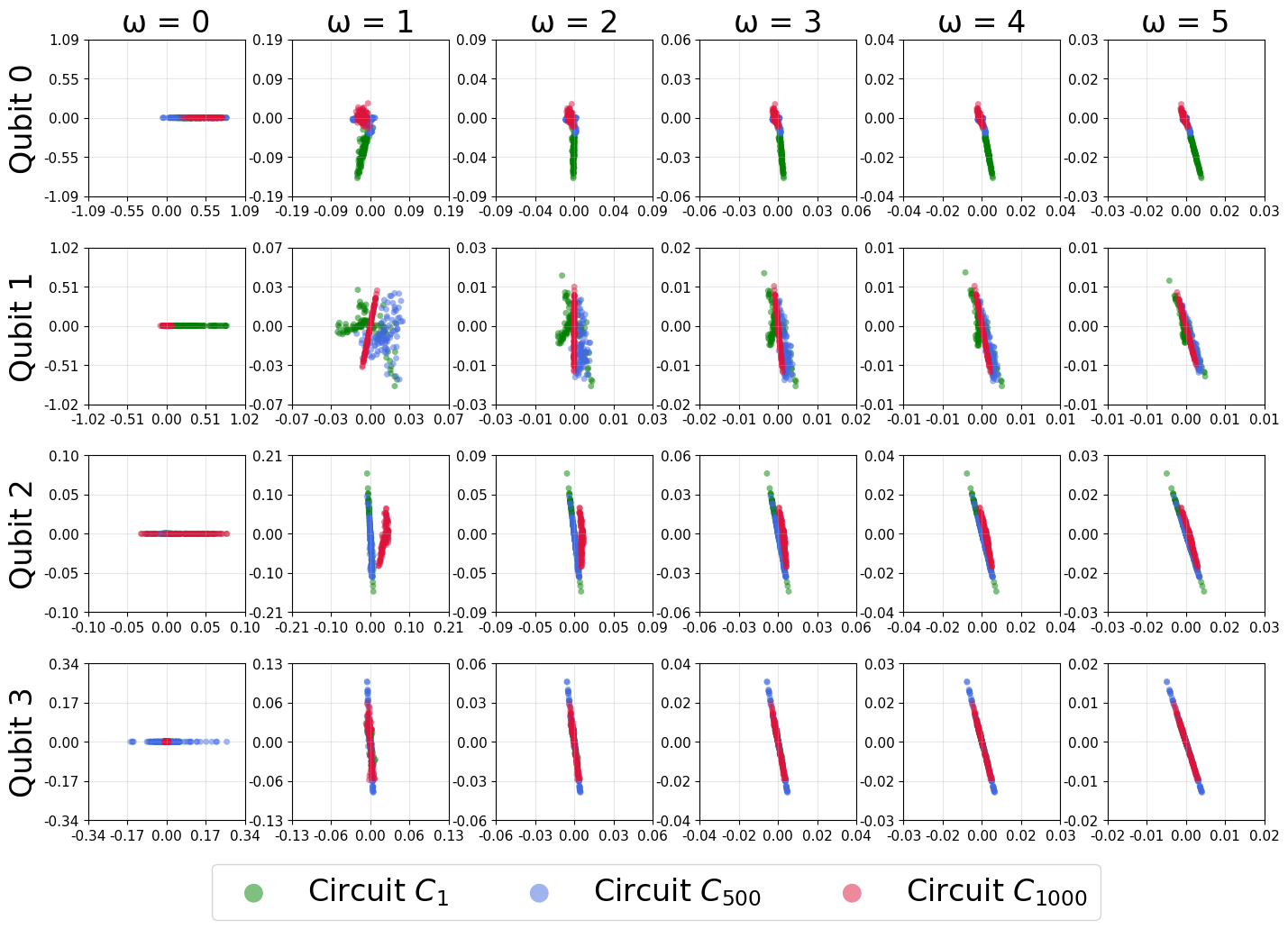}
    \caption{Distribution of Fourier coefficients in the complex plane 
    for high ($C_1$), medium ($C_{500}$), and low ($C_{1000}$) performing encodings on BreastMNIST.}
    \label{fig:fourier}
\end{figure}

The distribution of Fourier coefficients for encoding circuits with high ($C_1$), medium ($C_{500}$), and low ($C_{1000}$) performance are shown in \Cref{fig:fourier}, where the subscript indicates the rank of the circuit among all searched encodings. We do not observe clear spectral differences between high-performing and low-performing encodings. For most qubits, the coefficients predominantly cluster along a line regardless of classification performance, indicating constrained phase relationships. The lowest-performing encoding exhibits this linear pattern most strongly, but the effect is weak and does not generalize when comparing across all encodings. We observe similar results for PneumoniaMNIST.

\subsubsection{Feature Map Diversity}
As described in \Cref{sec:performance_encodings} and depicted in \Cref{fig:Feature_maps}, there are indications that data encodings producing more diverse feature maps lead to better classification performance. To investigate this quantitatively, we employ the normalized effective rank metric introduced in \Cref{sec:effective_rank}.

We apply this metric to the feature maps produced by the quantum encoding layer. For a given encoding circuit and input image, the circuit produces a feature map $F(x_i)$ of dimensions $C \times H \times W$, where $C$ is the number of feature channels and $H$, $W$ are the spatial dimensions. We reshape this tensor into a matrix of size $C \times (H \cdot W)$ and compute its normalized effective rank. To estimate the feature map diversity of an encoding circuit, we compute the average normalized effective rank over a representative subset of images:
\begin{equation}
    \overline{\text{erank}} = \frac{1}{|S|} \sum_{x_i \in S} 
    \text{erank}_{\text{norm}}(F(x_i)),
\end{equation}
where $S$ is the set of sampled images. We sample 20 images from the training set using a balanced strategy, choosing five images per class, selected at evenly spaced intervals along the brightness distribution to ensure coverage of the full intensity range.

\begin{figure}[t]
        \centering
    \begin{subfigure}[b]{\columnwidth}
         \centerline{\includegraphics[width=0.70\columnwidth]{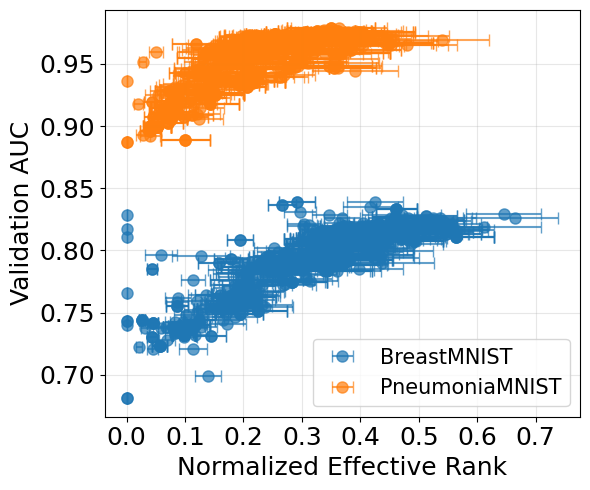}}
    \end{subfigure}

    \caption{Normalized effective rank against validation AUC for candidate encodings on both datasets. A higher effective rank indicates greater feature map diversity. The effective rank is computed on a subset of 20 images, with error bars indicating standard deviation. Pearson correlation: $r = 0.87$ (BreastMNIST) and $r = 0.82$ (PneumoniaMNIST).}
    \label{fig:feat_sim}
\end{figure}

We compute the average normalized effective rank for approximately 700 data encodings generated by the MCTS. To quantify performance, we use the best validation AUC achieved within 30 epochs for a single weight initialization. \Cref{fig:feat_sim} shows the average normalized effective rank plotted against the validation AUC for both the BreastMNIST and PneumoniaMNIST datasets. Error bars represent the standard deviation across the 20 sampled images per encoding. Both datasets exhibit a strong positive correlation between effective rank and validation AUC (Pearson correlation $r = 0.868$ for BreastMNIST and $r = 0.820$ for PneumoniaMNIST, both $p < 0.001$). This suggests that the effective rank may serve as a computationally efficient metric for identifying promising data encodings. 

However, examining the overall correlation alone can be misleading. To better understand where effective rank is most predictive, we split the data into quartiles by AUC and compute the correlation within each group. Both datasets exhibit a consistent pattern: correlation is strongest among poor-performing encodings (Q1: $r = 0.75$ for both datasets) and weakens progressively in the middle quartiles. For BreastMNIST, the correlation vanishes entirely among the best encodings (Q4: $r = -0.07$), indicating that effective rank provides no predictive power among the best-performing data encodings. PneumoniaMNIST retains moderate correlation in Q4 ($r = 0.43$), suggesting some level of predictive power even among high-quality encodings.

This correlation pattern has practical implications. While effective rank cannot reliably identify the very best encodings, it can reliably identify poor ones. We therefore propose using effective rank as a filter to eliminate unpromising candidates before full performance evaluation. This is substantially faster, as computing effective rank requires only a small subset of images rather than full training.

\begin{figure}[b]
        \centering
    \begin{subfigure}[t]{\columnwidth}
         \centerline{\includegraphics[width=0.76\columnwidth]{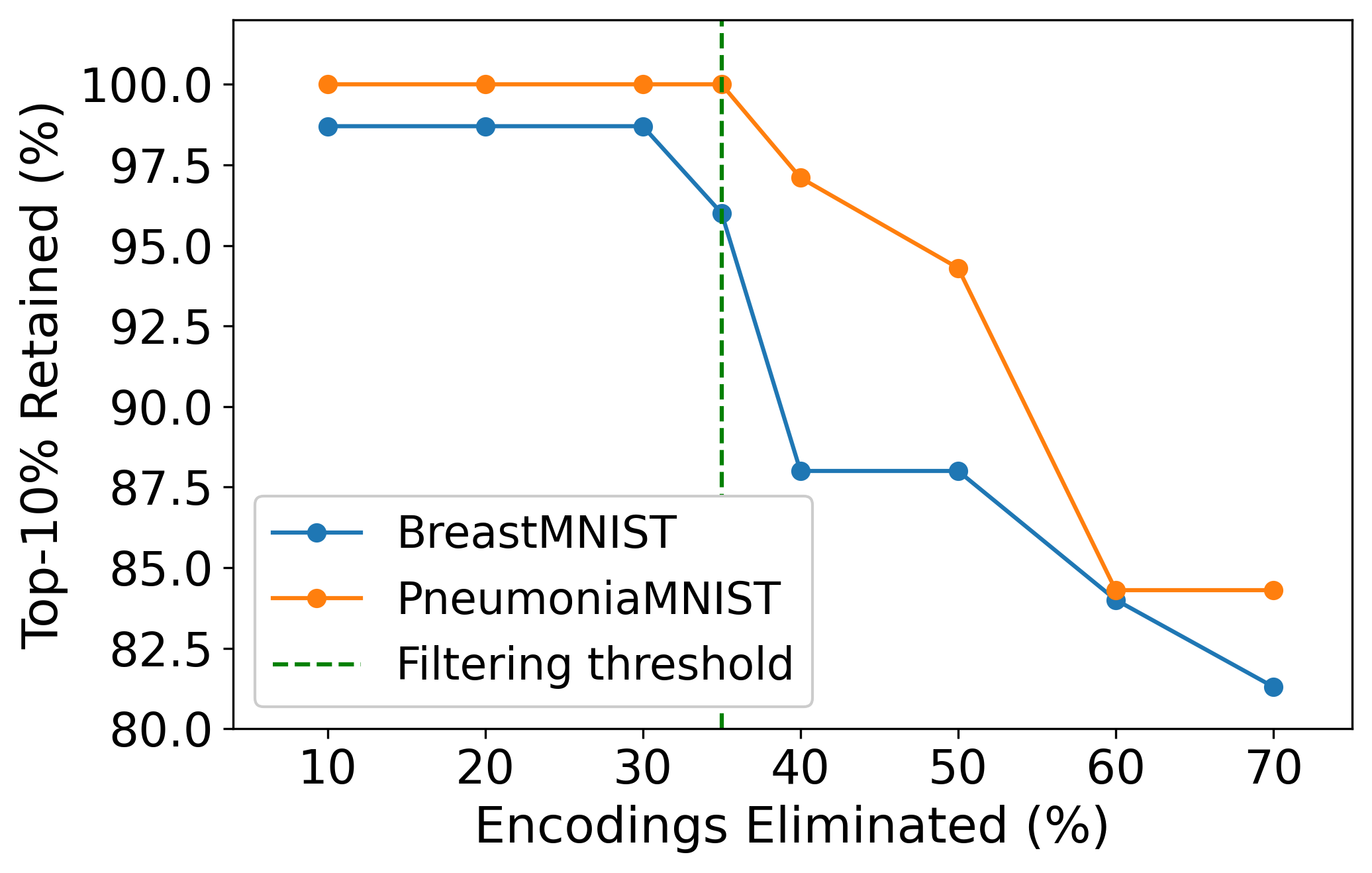}}
    \end{subfigure}

    \caption{Percentage of true top-10\% encodings (by AUC) retained after eliminating a bottom fraction (\%) by effective rank. The dashed line indicates the recommended filtering threshold at 35\%, which retains 100\% of top encodings for PneumoniaMNIST and 96\% for BreastMNIST.}
    \label{fig:filtering_strategy}
\end{figure}

To validate this filtering strategy, we simulate eliminating varying fractions of the lowest-ranked encodings by effective rank and measure what fraction of the top-10\% encodings would remain (\Cref{fig:filtering_strategy}). For PneumoniaMNIST, eliminating the bottom 35\% of candidates (those with effective rank $\leq 0.209$) preserves all of the top-10\% best candidates. For BreastMNIST, the same threshold (effective rank $\leq 0.283$) retains 96.0\% of the top-10\% encodings. Beyond 35\%, recall degrades more sharply, particularly for BreastMNIST, which drops to 88.0\% at the 40\% threshold, making more aggressive filtering inadvisable.

Based on these results, effective rank can serve as a computationally cheap filter to accelerate the search process. We propose integrating it as an early rejection criterion within the MCTS: when a new candidate is generated, its effective rank is computed first; if below a threshold, the encoding is rejected without full evaluation. The choice of threshold involves a trade-off between computational savings and the risk of discarding promising encodings. From our experiments, for PneumoniaMNIST all true top-10\% encodings had effective rank above 0.209, while for BreastMNIST 96\% had effective rank above 0.283. For new datasets, we recommend starting with a conservative threshold in the range of 0.20--0.25 and adjusting based on the observed distribution of effective ranks during the search.

To empirically validate this approach, we compared MCTS runs with and without effective rank pre-screening over approximately 400 search iterations on both datasets, using a threshold of 0.25. Computing effective rank requires approximately 7 seconds per candidate compared to over 300 seconds for full evaluation. For BreastMNIST, 35.9\% of candidates were rejected early, reducing total search time by approximately 35\%. The filtered approach achieved comparable results: AUC decreased marginally from 0.839 to 0.825, while accuracy (0.906 vs.\ 0.903) and loss (0.403 vs.\ 0.405) remained similar. For PneumoniaMNIST, the filtering rate was 50.3\%, reducing search time by approximately half. The filtered search achieved strong performance, with AUC reaching 1.0 (matching the unfiltered search), accuracy of 1.0 (vs.\ 0.979), and loss of 0.108 (vs.\ 0.280). While the variation between individual runs may partly reflect the stochastic nature of the search process, these results indicate that effective rank filtering substantially reduces computational cost without degrading the quality of discovered encodings. By rejecting low-quality candidates early, the search can focus computational resources on more promising regions of the encoding space. We note that the filtering rate is expected to decrease as the search progresses: once the MCTS concentrates on high-quality regions, most candidates will pass the effective rank threshold and require full evaluation. The primary computational savings therefore occur during the early exploration phase, where poor candidates are most prevalent.

%% file: sections/05_conclusion.tex
\section{Discussion and Conclusion} \label{sec: conclusion}
We presented an approach for discovering data encoding circuits for QCCNNs using MCTS. By removing the variational component entirely, our architecture addresses whether hybrid model performance stems primarily from data encoding or variational optimization. Our experiments on two medical imaging datasets demonstrate that automatically discovered encodings outperform commonly used encoding strategies and achieve competitive performance compared to classical baselines. Our metric analysis reveals that feature map diversity correlates strongly with performance and can serve as an early rejection criterion to accelerate the search process, while entanglement capability and Fourier properties show weak correlation.

Several limitations should be acknowledged. All simulations are performed in a noiseless environment, real quantum hardware may affect encoding performance. Our experiments are limited to binary classification on small datasets with minimal classical components. Importantly, our metrics measure representational richness and do not certify that the circuits provide a genuinely quantum computational advantage. Indeed, the MCTS algorithm shows no clear bias towards non-simulatable circuits, as seen with the entanglement capability, meaning that some discovered encodings may fall within regimes accessible to classical simulation methods such as tensor networks.

Several directions for future work emerge from these findings. First, extending the evaluation to additional datasets, including multiclass problems, three-dimensional medical images, and higher-resolution data with dedicated encoding searches, would test generalizability. Second, exploring more expressive classical architectures as the trainable component may reveal complementary strengths between classical and quantum feature extraction. Third, validation on real quantum hardware and noise-aware search strategies would assess practical applicability. Finally, the question of simulatability warrants further investigation: quantum-inspired classical methods could probe discovered circuits at larger scales, while modified search constraints might reveal conditions under which non-simulatable circuits emerge as more performant.